\documentclass[12pt]{iopart}
\usepackage{epsfig}

\begin{document}
\title[Quasi-Universal Dipolar Scattering]{Quasi-Universal Dipolar Scattering in
Cold and Ultracold Gases}
\author{J. L. Bohn${}^1$, M. Cavagnero${}^2$, and C. Ticknor${}^3$}
\address{${}^1$ JILA, NIST and Department of Physics, University of Colorado, Boulder, Colorado 80309-0440, USA \\ ${}^2$ Department of Physics and Astronomy, University of Kentucky, Lexington, KY 40506-0055 USA\\${}^3$ ARC Centre of Excellence for Quantum-Atom Optics and Centre for Atom Optics and Ultrafast Spectroscopy, Swinburne University of Technology, Hawthorn, Vic 3122, Australia}
\ead{${}^1$bohn@murphy.colorado.edu; ${}^2$mike@pa.uky.edu; ${}^3$cticknor@swin.edu.au}
\begin{abstract}
We investigate the scattering cross section of aligned dipolar
molecules in low-temperature gases. Over a wide range of collision
energies relevant to contemporary experiments, the cross section
declines in inverse proportion to the collision speed, and is given
nearly exactly by a simple semiclassical formula.  At yet lower
energies, the cross section becomes independent of energy, and is
reproduced within the Born approximation to within corrections due
to the s-wave scattering length.  While these behaviors are
universal for all polar molecules, nevertheless interesting
deviations from universality are expected to occur in the
intermediate energy range.
\end{abstract}

\maketitle
\section{Introduction}

The energy dependence of scattering cross sections for atoms at
ultralow collision energies is very simple.  Either the cross
section is nearly independent of energy, for distinguishable
particles or identical bosons; or else the cross section vanishes
altogether, for identical fermions.  This behavior emerges in the
limit where the deBroglie wavelength exceeds any natural length
scale of the interatomic interaction, and the scattering is
characterized by a single quantity, the s-wave scattering length $a$
(alternatively, the p-wave scattering volume $V_p$ for fermions).
Although $a$ dominates the threshold scattering, nevertheless its
value is not immediately obvious from the interaction potential, and
must be determined painstakingly from experiments.

By contrast, for low-energy collisions between polarized dipolar
molecules, the near-threshold scattering {\it is} often
approximately determined directly by parameters of the interaction
potential.  The interaction between two molecules of reduced mass
$M$ and dipole moment $\mu$ is characterized by a dipole length,
given by $D =M \mu^2 / \hbar^2$. \footnote{$D$ is determined,
roughly, by equating a typical centrifugal energy, $\hbar^2/MD^2$,
to a typical dipolar energy, $\mu^2/D^3$.  This is the same
reasoning that leads to the definition of the Bohr radius, by
equating centrifugal and Coulomb energies for hydrogen.} Under a
wide variety of circumstances, to be discussed in this paper, $D$ is
the dominant length scale and sets the threshold cross section,
i.e., $\sigma \sim D^2$. Moreover, this circumstance holds for
identical fermions as well as bosons. In this sense, the scattering
of two dipoles is nearly universal at threshold, apart from possible
modifications arising from s-wave scattering.

For realistic collisions in present-day experiments, however,
collision energies are not always in this threshold region. When the
threshold region is left behind, there is significant numerical
evidence to suggest that a universal behavior still emerges, and
that the cross section scales as $\sigma \sim D/K$, where $K$ is the
wave number of the relative motion \cite{CTuniversal}. The switch
between the two types of behavior corresponds roughly to the natural
energy scale of the dipolar interaction, $E_D = \mu^2/D^3
=\hbar^6/M^3 \mu^4$. Below this energy nonzero partial waves
contribute only perturbatively, and only at large intermolecular
separation; whereas at higher energies, many partial waves
contribute and the scattering is semiclassical.

These two behaviors may thus be said to define the operational limit
between ``cold'' and ``ultracold'' regimes of scattering for
dipoles.  We propose that the onset of cold collisions occurs when
the temperature of the gas passes well below the molecule's
rotational constant $B_e$ (or else its $\Lambda$-doublet splitting)
so that the orientational degrees of freedom freeze out.  These
temperatures are typically in the mK-K range.  This cold collision
regime, in which semiclassical scattering occurs, persists until the
temperature gets as low as $E_D$ (typically nK-$\mu$K temperatures
for dipolar molecules). Temperatures below $E_D$ define the
ultracold regime, where true threshold scattering is apparent.
Actual values of these temperatures depend strongly on the species
considered and on the applied electric field.

Our goal in this article is to make these ideas precise. We will
illustrate the universal behavior of dipole-dipole scattering at low
temperatures in the two regimes, and, more importantly, we will see
the circumstances under which this universality fails. Figure 1
shows the basic elements of universality for dipole-dipole
scattering. This figure plots the total scattering cross section
$\sigma$, averaged over all incident directions, versus collision
energy. Both quantities are presented in terms of the ``natural''
units given above.  The black curve is a complete numerical
close-coupling calculation. At low energies, $E<E_D$, $\sigma$
approaches a constant value that is well-approximated using the Born
approximation (blue).  At higher energies $E>E_D$, $\sigma$ falls
off as $1/\sqrt{E}$, and is given by the semiclassical eikonal
approximation (red).  Both approximations will be derived below.

\begin{figure}
\includegraphics[width=0.9\textwidth] {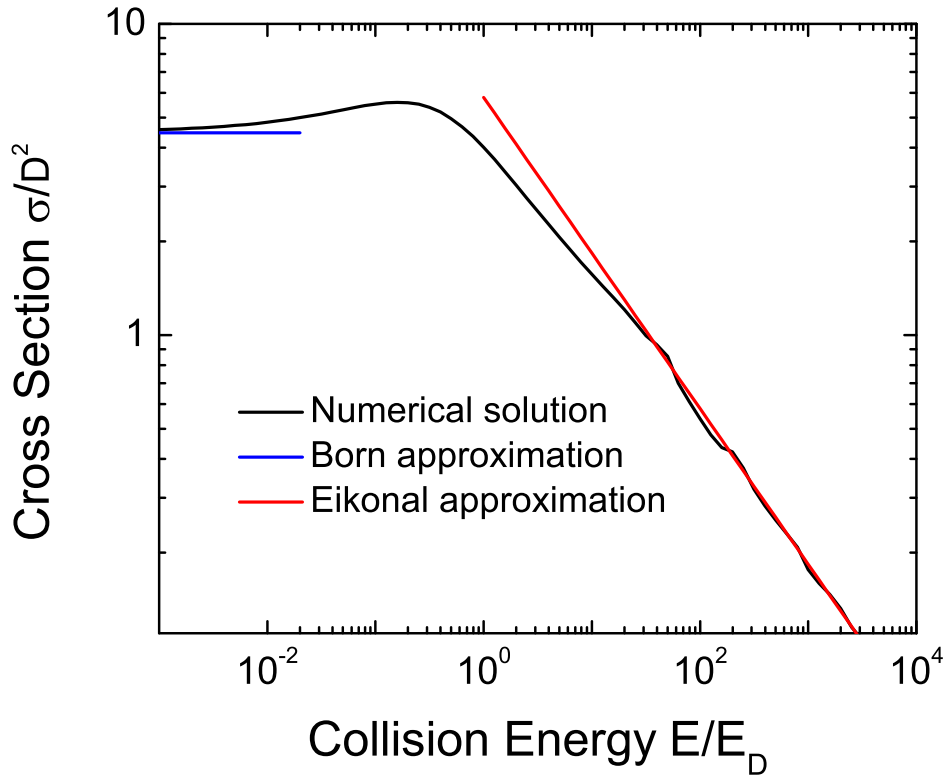} \caption{The
total scattering cross section $\sigma$, averaged over all incident
directions, for two distinguishable polarized dipoles. In the
low-energy limit, $\sigma$ reduces to the Born approximation result,
while at high energies it is given by a semiclassical eikonal
approximation. Note, however, that boundary conditions were
carefully chosen in this calculations so that the s-wave scattering
length vanishes.}
\end{figure}

In this article, we present alternative close-coupling calculations that
reveal deviations from this universal behavior.  For example, at
threshold, the cross section can deviate substantially from the Born
result. This occurs when threshold resonances in s-wave scattering
dramatically increase the cross section above the $D^2$
value~\cite{Ticknor05_PRA,Bohn05_ICLP,mike,mike2}.  More
interestingly, in the intermediate energy regime $E \approx E_D$
where neither approximation holds, our close-coupling calculations
show that the behavior of cross section versus energy depends on
details of the physics on length scales small compared to $D$. For
this reason, it is conceivable that elastic scattering experiments
that probe deviations from universality will be an important tool in
unraveling information on the short-range physics of close
encounters. Finally, we also discuss the angle-dependence of cold
collisions, in which the incident direction of the collision
partners is varied with respect to the polarization axis; this is a
measure of the importance of the anisotropy of the dipole-dipole
interaction. These results should serve as simple and accurate
guidelines to low-energy collision cross sections needed to
understand scattering or thermal equilibration in contemporary
experiments.

\section{Formulation of the problem}

For a pair of dipoles with reduced mass $M$ and polarized in the
${\hat z}$ direction by an external field, the two-body
Schr\"{o}dinger equation reads
\begin{eqnarray}\label{raw_Schrodinger}
\left[ - {\hbar ^2 \over 2 M} \bigtriangledown ^2 + \langle \mu_1
\rangle \langle \mu_2 \rangle {1 - 3\cos \theta \over R^3} + V_{SR}
\right] \psi = E \psi.
\end{eqnarray}
Here ${\vec R} = (R, \theta, \phi)$ is the relative displacement of
the two dipoles, and $\langle \mu_1 \rangle$ and $\langle \mu_2
\rangle$ are their induced dipole moments.  These dipoles are field
dependent, and their values are set by the field and the internal
structure of the dipole.  In the present context, we will take them
as fixed quantities, and observe their influence on the scattering.
By ignoring internal molecular structure, we are in effect modeling
molecules trapped in their absolute ground state, as in an optical
dipole trap.  In particular, we assume that inelastic scattering is
disallowed, a topic taken up in the next article of this issue~\cite{Newell}.

The potential term $V_{SR}$ represents all the short-range physics,
i.e., the potential energy surface of chemical significance when the
dipoles are close together.  This detail will be specific to each
pair of collision partners considered.  To simplify the discussion,
we will replace the complex details of $V_{SR}$ by imposing a
boundary condition at a fixed interparticle separation $R_0$.  We
will assert that the wave function $\psi$ vanishes uniformly on this
boundary, although alternative boundary conditions are certainly
possible and indeed desirable \cite{Ticknor05_PRA,Bohn05_ICLP}.
Having made this approximation, we ignore $V_{SR}$ in the following.

The resulting Shr\"{o}dinger equation
then admits a natural length
scale $D = M \langle \mu_1  \rangle \langle \mu_2 \rangle /
\hbar^2$, and a natural energy scale $E_D = \hbar^6 / M^3 \langle \mu_1
\rangle^2 \langle \mu_2 \rangle^2$.  By recasting (\ref{raw_Schrodinger})
in the scaled units
$r=R/D$, $\epsilon = E/E_D$, and by ignoring $V_{SR}$, we arrive at
the universal Schr\"{o}dinger equation (assuming that the molecules
have the same orientation relative to the field axis)
\begin{eqnarray}\label{scaled_Schrodinger}
\left[ - {1 \over 2} \bigtriangledown^2 - {2 C_{20} \over r^3}
\right] \psi = \epsilon \psi,
\end{eqnarray}
where $C_{20}(\theta,\phi) = (3 \cos^2 \theta - 1)/2$ is the
standard reduced spherical harmonic \cite{brink}.  Although the {\it
equation} (\ref{scaled_Schrodinger}) has a universal form,
nevertheless its {\it solutions} may depend on details of the
short-range physics, represented here by the cutoff radius $r_0 =
R_0/D$.  Our main objective is to explore circumstances under which
solutions are universal, i.e., independent of $r_0$, under the
conditions of modern cold molecule experiments.

As a preliminary argument in this direction, let us consider the
relative sizes of the two characteristic lengths $D$ and $R_0$.  For
most atomic and molecular species, the size scale $R_0$ below which
short-range physics can matter is on the order of the van der Waals
length, which is typically $\sim 100$ $ a_0$, which is also the
distance at which the internal fields generated by the dipoles are comparable to the
applied field (at least for typical laboratory field strengths).  By contrast, for
molecules with typical 1 Debye dipole moments, $D$ will be orders of
magnitude larger.  For example, in the OH radical, with $\mu =1.68$
Debye,  $D_{\rm OH} = 6770$ $ a_0$;  for the representative alkali
dimer KRb, with $\mu = 0.566$ Debye, $D_{\rm KRb} = 5740$ $a_0$; and
for the highly polar LiCs molecule with $\mu =5.5$ Debye, we find
$D_{\rm LiCs} = 6 \times 10^5$ $a_0$.  Correspondingly, the
characteristic energies are low: $E_{D,{\rm OH}} = 445$ nK, while
$E_{D,{\rm KRb}} = 83$ nK, and $E_{D,{\rm LiCs}} = 7$ pK. In making
these estimates, we assume the molecules are completely polarized;
$D$ gets shorter, and $E_D$ higher, if they are only partially
polarized. \footnote{For species whose {\it magnetic} dipole moment
mediates the interaction, these scales can be quite different.  For
atomic chromium, for instance, one finds $D_{Cr}=27$ $a_0$, far
smaller than its natural scattering length. A substantial body of
literature now treats the complete details of the Cr-Cr interaction
potential, for example
\cite{Werner05_PRL,Pavlovic05_PRA,Stuhler07_JMO}.}

The finite size of $R_0$ can destroy universality in three ways.
First, at low energies, we will see that scattering cross sections
are of order $D^2$.  This would be spoiled by the geometric cross
section $\propto R_0^2$, if $R_0$ were comparable to, or larger
than, the dipole length $D$ itself.  However, we have just argued
that this is usually not the case for polar molecules.  Second,
there is the possibility that the s-wave scattering length $a$ is
larger than $D$, and this would also alter the universal result at
low energy. Indeed, this very idea has been invoked as a means of
tuning the interaction between dipolar molecules
\cite{You,Ronen,Derevianko,Wang,Blume} or else as a tool for probing
details of these interactions \cite{Ticknor05_PRA}.  Often, the
effect of the scattering length is non-negligible, as we will see
below.

The third instance where $R_0$ may matter is in the extreme high energy
limit, where the universal cross section falls as $D/K$, where $K$
is the wave number.  In this case the cross sections will again tend
to the geometrical $\sim R_0^2$ for our artificially imposed hard
wall, and will dominate at energies where $R_0^2 > D/K$, which
translates to about 4.5 Kelvin in OH, well above Stark decelerator
energies, and at which point other degrees of freedom of the molecule
are relevant.  Thus non-universal behavior may not be a concern at
``high'' energies either, from the standpoint of current
experimental investigations.

In this article we compute scattering cross sections for dipoles
over a wide range of collision energies.  We do this in three ways:
1) a fully numerical close-coupling expansion of the wave function
in partial waves; 2) a Born approximation that exhibits the correct
universal behavior ($\sigma \sim D^2$) in the ultracold limit
($E<E_D$); and 3) a semiclassical eikonal approximation that
exhibits the correct universal behavior ($\sigma \sim D/K$) in the
cold collision regime ($E>E_D$). We briefly describe these methods
in the following subsections.

\subsection{Close-coupling formalism}

We expand the total wave function into partial waves in the
conventional way:
\begin{eqnarray}\label{partial_wave_expansion}
\psi(r,\theta, \phi) = {1 \over r} \sum_{lm} F_{lm}(r) Y_{lm}
(\theta ,\phi),
\end{eqnarray}
where the radial functions $F_{lm}$ satisfy a set of coupled-channel
Schr\"{o}dinger equations
\begin{eqnarray}\label{coupled_Schrodinger}
-{1 \over 2} {d^2 F_{lm} \over dr^2} + {l(l+1) \over 2 r^2}F_{lm} -
{2 \over r^3} \sum_{l^{\prime}} C_{l l^{\prime}}^{(m)} F_{l^\prime
m} = \epsilon F_{lm},
\end{eqnarray}
and the coupling matrix element is given by \cite{brink}
\begin{eqnarray}\label{coupling_matrix}
C_{l l^{\prime}}^{(m)} &=& \langle lm | C_{20} | l^{\prime} m
\rangle \\
&=& (-1)^m \sqrt{ (2l+1)(2l^{\prime}+1) } \left(
\begin{array}{ccc}
l & 2 & l^{\prime} \\
-m & 0 & m \end{array} \right) \left( \begin{array}{ccc} l & 2 &
l^{\prime} \\ 0 & 0 & 0 \end{array}\right) \nonumber.
\end{eqnarray}
Owing to the cylindrical symmetry of the Hamiltonian, the angular
momentum projection $m$ is a good quantum number, and we can solve a
separate set of coupled equations for each value of $m$.  However,
the boundary conditions of the wave function may not respect this
symmetry, i.e., the incident wave could arrive from any direction,
not just along the symmetry ($z$) axis. Thus we consider a complete
sum over $m$ in the wave function (\ref{partial_wave_expansion}).
Similarly, symmetries of the 3-$j$ symbols in
(\ref{coupling_matrix}) guarantee that each partial wave $l$ is
coupled only to the partial waves $l^{\prime}=l$, $l \pm 2$ by the
dipole interaction. Thus we can consider even partial waves
separately from odd partial waves, and will do so in the following.
For identical particles in the same internal state, these situations
correspond to bosons and fermions, respectively.

The equations (\ref{coupled_Schrodinger}) admit as many linearly
independent solutions as there are channels $(lm)$.  Individually,
they are defined by the boundary conditions (for each $m$)
\begin{eqnarray}\label{boundary_conditions}
F_{lm}^{l^{\prime} m}(r=r_0) &=& 0 \nonumber \\
F_{lm}^{l^{\prime} m} (r \rightarrow \infty) &=& \delta_{l
l^{\prime}}e^{-i(kr - l^{\prime} \pi /2)} - S_{l
l^{\prime}}^{(m)}e^{i(kr - l^{\prime} \pi /2)},
\end{eqnarray}
where $k = \sqrt{ 2 \epsilon}=DK$ is the wave number in dipole units.
These scattering boundary conditions serve to define the scattering
matrix $S_{l l^{\prime}}^{(m)}$.  From this matrix one can construct
the scattering amplitude describing scattering in direction ${\hat
k}_f = (\theta, \phi)$ from an incident direction ${\hat k}_i$
\cite{Mott,Bohn00_PRA}:
\begin{eqnarray}\label{scattering_amplitude}
f({\hat k}_i,{\hat k}_f) = -{2 \pi \over k} \sum_{l l^{\prime} m}
i^l Y_{lm}^*({\hat k}_i) T_{l l^{\prime}}^{(m)}
i^{-l^{\prime}}Y_{l^{\prime} m}({\hat k}_f),
\end{eqnarray}
in terms of the $T$ matrix, $T = i(S-I)$.  By integrating over the
final directions, we arrive at the total cross section for dipoles
incident along ${\hat k}_i$:
\begin{eqnarray}\label{sigma_tot}
\frac{\sigma_{{\rm tot}}({\hat k}_i)}{D^2} &=& \int d \phi d(\cos \theta) |f|^2
\nonumber \\
&=& {4 \pi \over k} \Im  f({\hat k}_i,{\hat k}_i).
\end{eqnarray}
This last line is the familiar optical theorem result.  This is the
type of cross section that can be measured in cold beam experiments,
where, say, one species is trapped and the other is incident on the
trap from the terminus of a Stark decelerator.  If the molecules are
magnetically trapped, then an electric field can be applied an at
arbitrary angle relative to the collision axis \cite{Sawyer07_PRL}.

Finally, the total cross section, integrated over an assumed
isotropic distribution of initial directions, is
\begin{eqnarray}
\frac{\sigma}{D^2} &=& \int d {\hat k}_i \frac{\sigma_{{\rm tot}}({\hat k}_i)}{D^2}
\nonumber
\\ &=& {\pi \over k^2} \sum_{l l^{\prime} m} |T_{l l^{\prime}}^{(m)}|^2.
\end{eqnarray}
This cross section is more relevant to {\it in situ} collisions in a
trap, which serve to re-thermalize the gas and provide evaporative
cooling.

Numerical solutions to the coupled-channel equations
(\ref{coupled_Schrodinger}) are determined using a variable stepsize
version of Johnson's algorithm \cite{johnson}.  To ensure
convergence of total cross sections, we include partial waves up to
$l \sim 100$ at the highest collision energies of $\epsilon = 10^4$.
Vice versa, at the lowest collision energies we can get away with
partial waves up to $l \sim 30$, but must apply the boundary
conditions (\ref{boundary_conditions}) as far out as $r = 20,000$.

\subsection{Born approximation}

At the lowest collision energies, the Wigner threshold laws are well
known to be different for dipolar interactions than for, say, van
der Waals interactions.  The elastic scattering phase shift
$\delta_l$ in partial wave $l>0$, due to a potential with $1/r^s$
long-range behavior, is \cite{Sadeghpour00_JPB}
\begin{eqnarray}\label{general_Wigner}
\tan \delta_l \sim A k^{2l+1} + B k^{s-2},
\end{eqnarray}
for some constants $A$ and $B$ that depend on short-range details.
The first term in (\ref{general_Wigner}) arises from the action of
the short range potential, while the second is due to purely
long-range scattering {\it outside} the centrifugal barrier.  For
the van der Waals potential ($s=6$), both contributions go to zero
faster than $\sim k$ at zero energy, and do not contribute to the
threshold cross section. However, for the dipole-dipole interaction
($s=3$), the second term is $\sim k$ for all partial waves.  The
contribution to the cross section, $\propto \sin^2 \delta_l/k^2$, is
then independent of energy in all partial waves, and this cross
section arises from long-range scattering.

This circumstance leads to the applicability of the Born
approximation in threshold scattering of dipoles
\cite{Deb,Derevianko,Wang,Li01_PRA,Avdeenkov05_PRA,Blume}.  At ever
lower collision energy, scattering occurs at ever larger values of
$r$ outside the barrier. But at long range the dipole-dipole
interaction $\propto 1/r^3$ is weak, and the perturbative Born
approximation is applicable.  These remarks do not apply, however,
to s-wave scattering, where there is no barrier.

The Born approximation for the scattering amplitude reads
\begin{eqnarray}
f({\hat k}_i,{\hat k}_f) = -{1\over 2 \pi} \int d^3 r e^{-i{\vec
k}_f \cdot {\vec r}} V_d({\vec r}) e^{i{\vec k}_i \cdot {\vec r}}.
\end{eqnarray}
Replacing each plane wave by its standard partial wave
expansion and re-arranging yields
\begin{eqnarray}
f({\hat k}_i,{\hat k}_f) = -{1 \over 2 \pi} \int d^3r \left( - {2
\over r^3} \right) C_{20}({\hat r}) && 4 \pi
\sum_{l^{\prime}m^{\prime}} i^{-l^{\prime}} Y_{l^{
\prime}m^{\prime}}({\hat k}_f)
Y_{l^{\prime}m^{\prime}}^*({\hat r}) j_{l^{\prime}}(kr) \nonumber \\
& \times & 4 \pi \sum_{lm} i^l Y_{lm}^*({\hat k}_i) Y_{lm}({\hat r})
j_l(kr).
\end{eqnarray}
Consolidating the integrals into radial and angular varieties, we
arrive at
\begin{eqnarray}\label{Born_f}
f({\hat k}_i,{\hat k}_f) = 2 \pi \sum_{l l^{\prime} m} i^l
Y_{lm}^*({\hat k}_i) C_{l l^{\prime}}^{(m)} \Gamma_{l l^{\prime}}
i^{-l^{\prime}} Y_{l^{ \prime}m^{\prime}}({\hat k}_f),
\end{eqnarray}
where $C_{l l^{\prime}}^{(m)}$ is the angular integral
(\ref{coupling_matrix}) defined above, and $\Gamma_{l l^{\prime}}$
is the radial integral
\begin{eqnarray}
\Gamma_{l l^{\prime}} &=& 8 \int_0^{\infty} r^2 dr {j_{l}(kr)
j_{l^{\prime}}(kr) \over r^3} \nonumber \\
&=&  {\pi \Gamma((l+l^{\prime})/2) \over \Gamma((-l+l^{\prime}+3)/2)
\Gamma((l+l^{\prime}+4)/2)
\Gamma((l-l^{\prime}+3)/2) }.  \nonumber \\
&=& \left\{ \begin{array}{l} {32 \over l(l+1)}, \;\;\;\;\;\;\;\;\; l^{\prime}=l \\
{32 \over 3(l+1)(l+2)}, \;\;\; l^{\prime} = l + 2 \end{array}
\right.
\end{eqnarray}
Comparing the Born result (\ref{Born_f}) with the expression
(\ref{scattering_amplitude}) (the factor of 8 is intended to
simplify this) identifies the $T$-matrix in the Born approximation
as
\begin{eqnarray}
T_{l l^{\prime}}^{(m),{\rm Born}} = - k C_{l
l^{\prime}}^{(m)}\Gamma_{l l^{\prime}}.
\end{eqnarray}

The Born approximation must be applied with a caveat.  For purely
s-wave scattering, where $l = l^{\prime} = 0$, the matrix element
$C_{00}^{(0)}$ vanishes, and so therefore does the $T$-matrix
element $T_{00}^{(0),{\rm Born}}$.  The Born approximation is
therefore mute on the question of s-wave scattering.  As argued
above, s-wave scattering, described by a scattering length $a$, is
part of the non-universal behavior of scattering anyway.  To produce
realistic scattering results, it is possible to supplement $T^{{\rm
Born}}$ with an empirical s-wave contribution, which is determined
from the full close-coupling calculations
\cite{You,Derevianko,Blume, Wang}.

Summarizing all these results, the threshold cross section in the
Born approximation, averaged over incident directions, is given by
the incoherent sum
\begin{eqnarray}
\frac{\sigma_{{\rm Born}}}{D^2} = {\pi \over k^2} \sum_{ll^{\prime}m}
|T_{ll^{\prime}}^{(m),{\rm Born}}|^2
\end{eqnarray}
Evaluating the sums, the even and odd partial wave cross sections at
threshold can be given as
\begin{eqnarray}\label{born_cross_sections}
\sigma_{{\rm Born}}^e = && 1.117 D^2  + 4 \pi a^2 \nonumber \\
\sigma_{{\rm Born}}^o=&& 3.351 D^2 ,
\end{eqnarray}
where $4 \pi a^2$ allows for the existence of a scattering length
$a$, which is not determined in the Born approximation.  For
identical particles, the cross sections (\ref{born_cross_sections})
must be multiplied by 2, as usual; for distinguishable particles,
both even and odd partial waves are possible, and these cross
sections are to be added.

The low-energy limits (\ref{born_cross_sections}) were independently
verified in near-threshold calculations using a coupled-channel
adiabatic representation~\cite{mike,mike2}, which also illustrated
how threshold angular distributions are affected by the competetion
between long-range (Born) coupling and s-wave scattering
resonances, as discussed in Section 4.

\subsection{Eikonal approximation}

At sufficiently high energies, a semi-classical analysis yields the
simple $D/K$ scaling of the cross section. Note, in reference to
Fig. (1), that the deBroglie wavelength becomes smaller than the
natural dipole length scale, $2\pi/K<D$, when $E>2\pi^2E_D$. This
marks the onset of semi-classical scattering. This semi-classical
onset can lie at $\mu K$ temperatures, or even colder, owing to the
large dipole length scale and therefore the small value of $E_D$.
Many partial waves contribute to the scattering amplitude in the
semi-classical regime, and differential cross-sections are
increasingly concentrated in the forward direction.

The eikonal method was long ago developed to find approximate
scattering solutions of wave equations such as
(\ref{scaled_Schrodinger}) valid in the semi-classical or ray-optics
limit, in which the potential is assumed to vary little on the scale
of the wavelength. A derivation of the eikonal wavefunction will not
be given here, as it can be found in familiar texts
\cite{Bransden-Joachain} and in a comprehensive review article by
Glauber \cite{Glauber}. Suffice it to say that a phase-amplitude
{\it ansatz}, coupled with the assumption of a slowly-varying
amplitude, leads directly to an approximate wavefunction
\begin{eqnarray}
\psi(\vec r) = e^{i\vec k_i\cdot\vec r}\exp\left[-{i\over k}\int^z V(b,\phi,z^{\prime}) dz^{\prime}\right]
\end{eqnarray}
Following Glauber, this wavefunction is expressed in cylindrical
coordinates with a new quantization (or $z$) axis aligned with the
average collision momentum, $\vec k_{\rm avg} = (\vec k_i+\vec
k_f)/2$. The cylindrical radius about this axis, $b$, can be
associated with a classical impact parameter: $\phi$ is the
azimuthal angle about the quantization axis. Due to the shift of
quantization axis away from the direction of the applied field,
$\vec {\cal E}$, we now note the field direction explicitly in the
potential $V(\vec r) = [1-3(\hat r\cdot\hat{\cal E})^2]/r^3$.

The two factors in the eikonal wavefunction are familiar in the
context  of the one-dimensional WKB method, where they coincide with
an expansion of the WKB phase $i\int dz~\sqrt{2(\epsilon-V)}\approx
ikz ~-~ (i/k)\int~dz~V$ to first order in $V/\epsilon$. Accordingly,
we anticipate that the eikonal method will be most accurate when the
incident energy is large compared to the magnitude of the
dipole-dipole interaction, a more stringent criterion than the
semi-classical constraint noted above.

The analysis of scattering amplitudes associated with the
approximate  eikonal wavefunction is simplified by a judicious
choice of coordinates. Note, in particular, that the momentum
transfer $\vec q = \vec k_i -\vec k_f$ is orthogonal to the
quantization axis defined by $\vec k_{\rm avg}$. For simplicity, we
define an $x$-axis along $\vec q$, in which case the $y$-axis is
orthogonal to the collision plane and lies along $\vec k_{\rm avg}
\times \vec q$. In this reference frame, the impact parameter is
written in vector form as $\vec b = b\cos(\phi)\hat x +
b\sin(\phi)\hat y$, and the relative displacement of the dipoles is
$\vec r = \vec b + z\hat k_{\rm avg}$.

Insertion of the eikonal wavefunction, valid where the potential is
non-negligible, into the integral equation for scattering leads to
the eikonal scattering amplitude
\begin{eqnarray}
f^{Ei}(\vec k_f,\vec k_i) = \frac{k}{2\pi i} \int~b ~ db~ d\phi
~e^{iqb\cos(\phi)} \left[e^{i\chi(\vec b)}-1\right]
\end{eqnarray}
where $k=\vert k_i\vert = \vert k_f\vert=\sqrt{2\epsilon}$, $q=\vert\vec q\vert = 2k\sin(\theta_s/2)$, and where the eikonal phase is
\begin{eqnarray}
\chi(\vec b) = -\frac{1}{k}
\int_{-\infty}^{\infty} dz^{\prime}~V(b,\phi,z^{\prime})
\end{eqnarray}

With the explicit form of the dipole-dipole potential
\begin{eqnarray}
V(b,\phi,z) = \frac{1}{(b^2+z^2)^{3/2}}
\left[
1-3\frac{(\vec b\cdot\hat {\cal E}+z\hat k_{\rm avg}\cdot\hat{\cal E})^2}
{b^2+z^2}\right]
\end{eqnarray}
the phase is readily evaluated; setting $\sigma = z/b$ one finds
\begin{eqnarray}
\chi &=& -\frac{1}{k b^2}\left[
\int_{-\infty}^{\infty}\frac{d\sigma}{(1+\sigma^2)^{3/2}}\right.\nonumber \\
&-&3(\hat b\cdot\hat{\cal E})^2
\int_{-\infty}^{\infty}\frac{d\sigma}{(1+\sigma^2)^{5/2}}\nonumber \\
&-&3(\hat k_{\rm avg}\cdot\hat{\cal E})^2
\left.\int_{-\infty}^{\infty}\frac{\sigma^2 d\sigma}{(1+\sigma^2)^{5/2}}\right]
\end{eqnarray}
The integrals are straightforward, giving
\begin{eqnarray}
\chi = -\frac{2}{k b^2}
\left[1-(\hat k_{\rm avg}\cdot\hat{\cal E})^2-2(\hat b\cdot\hat{\cal E})^2\right]
\end{eqnarray}
Referring the electric field to our coordinate
axes $(\hat x=\hat q, \hat y=\hat k_{\rm avg}\times\hat q,\hat k_{\rm avg})$
\begin{eqnarray}
\hat{\cal E} = \sin\alpha\cos\beta\hat x + \sin\alpha\sin\beta\hat y + \cos\alpha\hat k_{\rm avg}
\end{eqnarray}
the phase is simply
\begin{eqnarray}
\chi(b,\phi) = \frac{2}{k b^2}\sin^2\alpha \cos(2\phi-2\beta)
\end{eqnarray}

The eikonal amplitude now has the form
\begin{eqnarray}
f^{Ei} = \frac{k}{2\pi i}
\int~b~db~d\phi~e^{iqb\cos\phi}
\left[
\exp{\left\{i\frac{2}{k b^2}\sin^2\alpha\cos(2\phi-2\beta)\right\}}-1\right]
\end{eqnarray}
Explicit evaluation of the resulting integrals has proven quite
difficult. However, to extract total cross-sections, Glauber's general
proof of unitarity of the eikonal approximation~\cite{Glauber} permits
use of the
optical theorem, (\ref{sigma_tot}).

For forward scattering, $q=0$ and the first phase vanishes identically. Expressing the result in terms of the orbital angular momentum $l=kb$ gives
\begin{eqnarray}
f^{Ei}(\hat k_i,\hat k_i) = \frac{1}{2\pi i k}
\int~\ell~d\ell~d\phi~
\left[
\exp{\left\{i\frac{2k}{\ell^2}\sin^2\alpha\cos(2\phi-2\beta)\right\}}-1\right]
\end{eqnarray}
but note that this appears undetermined since $\hat q$ and, consequently, $\hat x$ and
$\hat y$ are not defined when $q=0$! In this limit, $\alpha = \arccos(\hat k\cdot\hat {\cal E})$ is well-defined, but $\beta = \arctan(\hat y\cdot\hat{\cal E}/\hat x\cdot\hat{\cal E})$
is not. Fortunately, it is easy to show that the azimuthal integral is independent
of $\beta$, with the result
\begin{eqnarray}
f^{Ei}(\hat k_i,\hat k_i) = \frac{1}{ik}\int~\ell~d\ell~\left[
J_0\left(\frac{2k}{\ell^2}\sin^2\alpha\right)-1\right]
\end{eqnarray}

From the optical theorem, the total cross section is then
\begin{eqnarray}
\frac{\sigma^{Ei}}{D^2} = \frac{4\pi}{k^2}\int_0^{\infty}~\ell~d\ell
\left[1-J_0\left(\frac{2k}{\ell^2}\sin^2\alpha\right)\right]
\end{eqnarray}
This result provides some insight into partial wave analysis in the semi-classical regime, which approximately separates into two regions: For $l<\sqrt{k}\sin(\alpha)$, the integrand is nearly linear in $l$, while for larger $l$ it declines steeply as $k^2 \sin^4(\alpha)/l^3$. Using these approximations, the integral evaluates to $k\sin^2(\alpha)$, the exact result given below.

More carefully, we set
\begin{eqnarray}
s = \frac{2k}{\ell^2}\sin^2\alpha
~~~,~~~\ell~d\ell = -k\sin^2\alpha \frac{ds}{s^2}
\end{eqnarray}
and so express the total eikonal cross section in terms of
a dimensionless integral
\begin{eqnarray}
\frac{\sigma^{Ei}}{D^2} = \frac{4\pi\sin^2\alpha}{k}
\int_0^{\infty}~\frac{ds}{s^2}\left[1-J_0(s)\right]
\end{eqnarray}
The integral evaluates to unity, with the result
\begin{eqnarray}\label{full_eikonal}
\frac{\sigma^{Ei}_{\rm tot}({\hat k}_i)}{D^2} =
\frac{4\pi}{k}\left[1-\left(\hat k_i\cdot\hat {\cal
E}\right)^2\right]
\end{eqnarray}
Remarkably, the cross section is identically zero when the direction of incidence is aligned with the field axis. Since, as will be discussed below, this semi-classical cross section accurately describes dipolar collisions in the temperature range currently accessible experimentally, there are a variety of observables which might test this angle-dependence of the total elastic cross section. While equilibrium properties of the gas would not be sensitive to the angle-dependence, non-equilibrium properties, such as transmission of fast dipoles through a trapped dipolar gas, would be expected to show strong dependence on the alignment of the beam with the field axis.

Averaged over incident directions within a confined gas, one then expects
\begin{eqnarray}\label{eikonal_sigma}
\frac{\sigma_{\rm Ei}}{D^2} = \frac{8\pi}{3k} = \frac{8\pi}{3KD}
\end{eqnarray}
which is the final result. As shown in Fig.~1, Glauber's method
yields not only the correct scaling, but quantitatively reproduces
the universal results discovered in close-coupling calculations
\cite{CTuniversal}.

The basic structure of this result was surmised by Gallagher
\cite{Gallagher} in a study of Rydberg-Rydberg collisions: from the
uncertainty principle, an interaction of energy $1/b^3$ and lasting
for a duration $b/k$, should satisfy $1/b^2k\sim 1$ in scaled units,
so that $\sigma\sim b^2 = 1/k$. Using an isotropic $-1/r^3$
interaction, DeMille \cite{DeMille} also applied the eikonal
approximation and determined a cross section $\sigma/D^2 =
2\pi^2/k$. Kajita's Fourier technique \cite{kajita} yields the
high-velocity result $\sigma/D^2 = 40\pi\sqrt{2}/3k$.

\section {The rise and fall of universal scattering}

Figure 1 has already presented the message of universality, in that
the Born and eikonal limits are achieved in the appropriate energy
ranges.  However, the calculations in this figure were carefully
selected to have zero scattering length, by choosing an appropriate
cutoff radius $r_0$.  By changing $r_0$, we are able to generate any
scattering length $a$. Experimentally, the value of $a$ can be
altered by changing the electric or magnetic field strength.
Changing $r_0$ could also change the scattering phase shift of any
other partial wave. However, this is only likely for those partial
waves whose centrifugal barrier lie below the collision energy,
because these partial waves are the only ones to probe physics at
the scale of $r_0$.

The importance of the centrifugal barriers in determining the range
of applicability of the low-energy (Born) scaling was emphasized in
\cite{mike,mike2}, through adiabatic calculations which converge
much more rapidly than close-coupling calculations at the lowest
energies. The adiabatic curves have pronounced barriers (in all but
the s-wave channel) separating repulsive centrifugal behavior at
large-$r$ from attractive dipolar behavior at small-$r$. The heights
of the lowest barriers (approximately coincident with $E_D$)
determine the range of energy over which the threshold behavior of
the cross section is approximately constant. They also suggest the
sensitivity to $r_0$ at intermediate energies, where incident flux
can surmount the barrier.

Figure 2 shows the effect of changing $r_0$ on the ``universal''
cross section from Fig. 1.  In Fig. 2a) is shown the result for even
partial waves.  The solid line is the $a=0$ result, and it amicably
reaches the universal Born and eikonal limits.  The other curves
employ values of $r_0$ that produce scattering lengths of $a = 0.1
D$ (red) and $a = -0.1 D$ (blue).  This change has made a
significant difference in the low-energy limit, where now the Born
approximation is merely a lower limit to the cross section. Notice
also that for $a<0$ the cross section initially decreases with
increasing energy, just as it does for alkali atoms
\cite{Bohn99_PRA}

However, at higher energies $E>E_D$, this change in $r_0$ has no
effect on $\sigma$.  One way to look at this is that the phase
shifts have changed for many partial waves, but because there are so
many of them added together to get the cross section, these changes
average out. Another point of view is that the semiclassical
scattering occurs at high impact parameter, and is thus indifferent
to what happens at $r = r_0$. In the intermediate energy range, the
change is still quite significant, since phase shifts are changing
for only those few partial waves that skip over their centrifugal
barriers.

\begin{figure}
\includegraphics[width=0.9\textwidth] {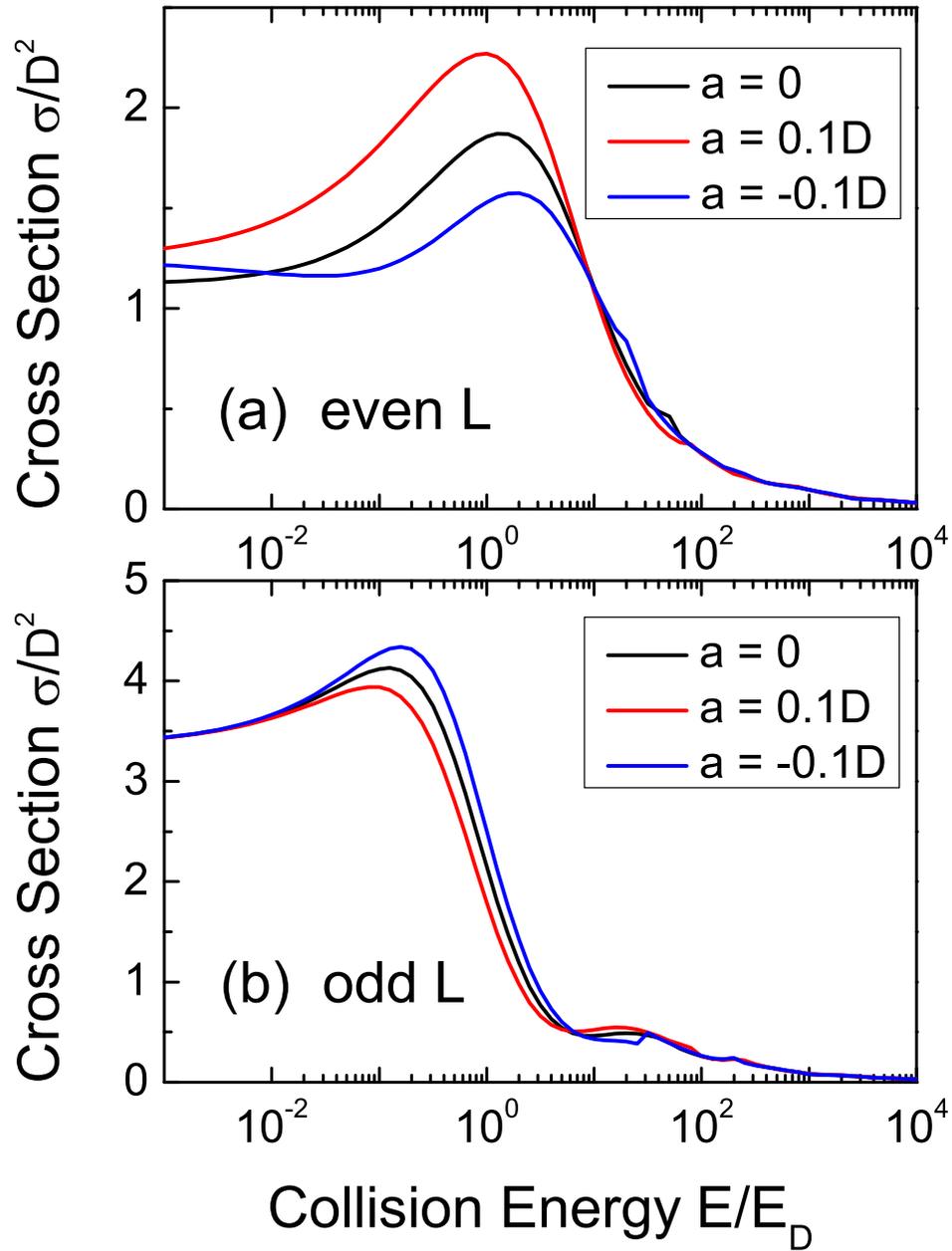} \caption{Cross sections
as in Fig. 1, but separated into contributions from even (a) or odd
(b) partial waves.  In each case, three different values of the
cutoff radius $r_0$ are chosen, corresponding to three different
s-wave scattering lengths $a$, as indicated in the legend.  In each
panel differences in the three curves demonstrate the breakdown of
universality.  Elastic scattering is thus much more universal for
odd partial waves (identical fermions) than for even partial waves
(identical bosons or distinguishable particles.}
\end{figure}

Figure 2b) shows the same circumstance, but for odd partial waves.
The same three values of $r_0$ are employed here, and so the three
curves are labeled by the scattering lengths from part a).  In this
case the cross sections always approach the Born limit, since there
is no aberrant s-wave scattering to derail them.  In the high-energy
limit, too, $\sigma$ is again insensitive to $r_0$.  It is only in
the intermediate energy range that a small deviation is seen.  This
suggests that for odd partial waves the behavior of the cross
section is indeed nearly universal. This would be true, for example,
in collisions of fermionic molecules (e.g., $^{40}$K$^{87}$Rb
\cite{gpm}) in identical hyperfine states.

To further emphasize the consequence of an s-wave scattering length,
Figure 3 reports the cross section $\sigma$ for the even partial
waves as a function of $r_0$, in the threshold limit, using $E/E_D =
10^{-3}$. The range of $r_0$ shown here corresponds to a complete
cycle of the scattering length from zero, through infinity, and back
to zero again. Consequently, the numerically evaluated cross section
shows a resonance, at which point the cross section is determined by
$a^2$, not $D^2$.  Even away from the resonance peak, the s-wave
contribution can significantly increase the cross section.  We are
thus led to conclude that universality at low-energies, in cases
where s-wave scattering is allowed, is similar to the universality
for atoms.  Namely, the form of the cross section (independent of
energy) is universal, but its value relies on a (field-dependent)
scattering length that must be determined empirically.  The Born
approximation does provide a useful lower limit, however.  For odd
partial waves, while resonances exist, they are shape resonances,
hence narrow at low energies and less likely to destroy universal
behavior.

The s-wave contribution to the Hamiltonian nominally vanishes, since
$C_{00}^{(0)}=0$.  However, it is not unreasonable that s-wave
scattering has a strong influence near threshold. To see this, we
evaluate an effective s-wave interaction at long range, via its
coupling to the $l=2$ partial wave, in second-order perturbation
theory (compare Ref. \cite{Avdeenkov02_PRA}):
\begin{eqnarray}
V_0(r) \approx -{ |2C_{02}^{(0)}/r^3|^2 \over l(l+1)/2r^2} = - {C_4
\over r^4},
\end{eqnarray}
where, in dipole units and using $l=2$, the coefficient is $C_4 =
4/3\sqrt{5}$.  This in turn leads to a characteristic length scale
for the s-wave interaction, analogous to the characteristic van der
Waals length,
\begin{eqnarray}
r_4 = \left( {2 \mu C_4 \over \hbar^2} \right)^{1/2} = \left( 2C_4
\right)^{1/2}.
\end{eqnarray}
In dipole units, this is $r_4 = 1.09 D$, comparable to the dipole
length itself.  Based on this consideration, it is perhaps not too
surprising that s-wave scattering plays a significant role.

\begin{figure}
\includegraphics[width=0.9\textwidth] {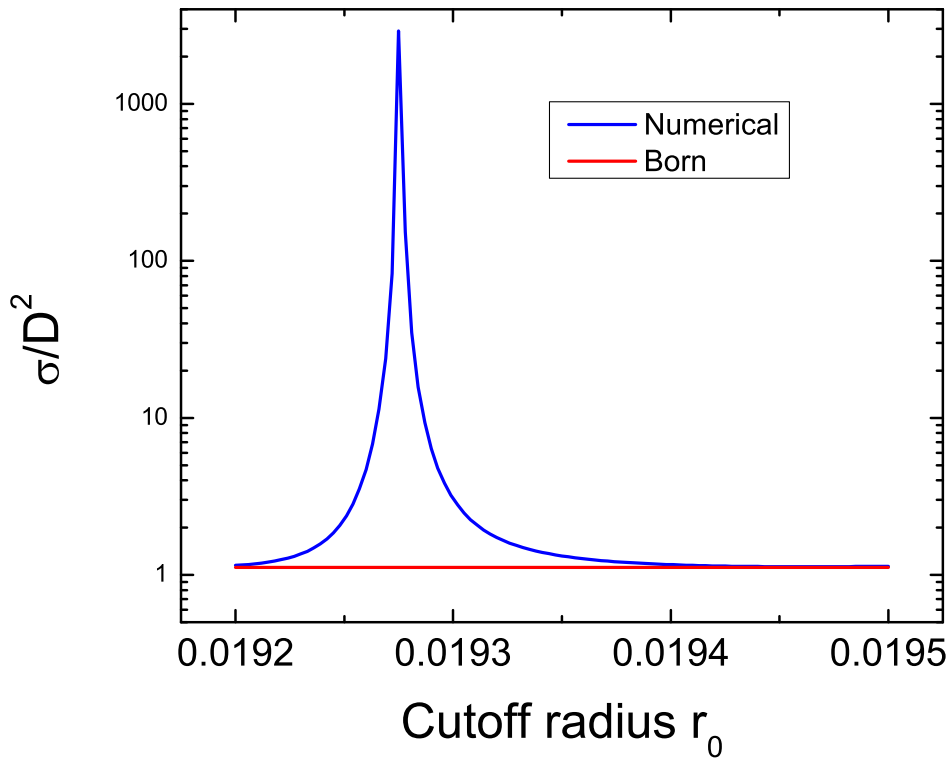}
\caption{Cross section $\sigma$ for various values of the cutoff
radius $r_0$.  The numerically determined cross section is always
larger then the Born approximation, sometimes significantly higher,
due to a large $s$-wave scattering length.}
\end{figure}

\section {Dependence on incident angle}

Thus far we have focused on the cross section as integrated over all
incident angles.  One of the interesting aspects of the
dipole-dipole interaction, however, is its anisotropy.  The cross
section $\sigma_{{\rm tot}}(\theta_i)$ may therefore depend on the
angle $\theta_i=\arccos(\hat k_i\cdot\hat {\cal E})$ of the incident
collision axis, with respect to the polarization direction.  This
cross section is easily calculated numerically from
(\ref{sigma_tot}), and also from the useful eikonal estimate
(\ref{full_eikonal}).

To show the utility of the eikonal expression, we present in Figure
4 $\sigma_{{\rm tot}}$ versus $\theta_i$, for a collision energy
$E/E_D=10^{4}$ where the eikonal approximation should be fairly
accurate. The total close-coupled cross section (black solid line)
is the sum of contributions from even and odd partial waves.  Note
that the contributions from these two sets of partial waves are
nearly equal here in the semiclassical limit where many partial
waves contribute, and effects of dipole-indistinguishability are
small. The angular distribution of each shows oscillations, but in
the sum, representing distinguishable particles, the angular
dependence is smooth.  Moreover, for angles where the collision axis
is orthogonal to the polarization axis, $\theta_i \approx \pi/2$,
the eikonal approximation (dotted line) is quite good.

A major deviation occurs, however, for dipoles aligned parallel to
the collision axis.  Here the eikonal result calls for vanishing
cross section, whereas the close coupling calculation yields a
non-zero cross section. For $\theta_i=0$, the incident wave
$e^{i\vec k_i\cdot\vec r}$ is invariant under rotations about the
field axis, and so contains only $m=0$ partial waves. It is not too
surprising that a semi-classical analysis will break down for
low-$m$ states. Furthermore, for $m=0$ states, the wavefunction is
large where the potential is strongest (near $r=0$), so the eikonal
assumption $V/\epsilon<<1$ is no longer valid. Most interesting
about this deviation from eikonal behavior, is the importance of
back-scattering from the strong potential in this geometry, as
indicated by the pronounced difference between even and odd partial
waves. Observations near $\theta_i=0$ will accordingly be most
sensitive to exchange scattering, and, presumably to short range
physics.

\begin{figure}
\includegraphics[width=0.9\textwidth] {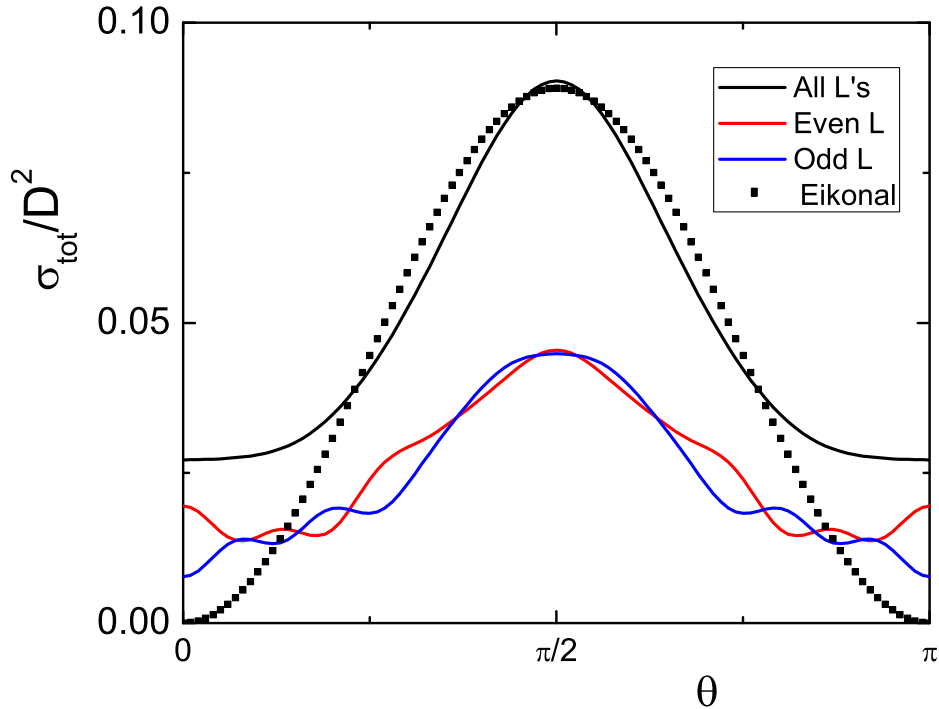} \caption{Dependence of
scattering cross sections $\sigma_{{\rm tot}}$ on the incident angle
$\theta_i$ between the collision axis and the polarization axis of
the dipoles.  This calculation is performed at a high collision
energy, $E/E_D = 10^{4}$.  The distribution is well-approximated
using the eikonal result (\ref{full_eikonal}), except when the collision and
polarization axes nearly coincide.}
\end{figure}

Cross section variations with the angle of incidence have also been
studied at low energies~\cite{mike,mike2}. When the $s$-wave
scattering length is negligible, a universal anisotropic
distribution is obtained, entirely due to long-range scattering.
However, when the $s$-wave scattering length dominates, near the
peak in Figure 3, a completely isotropic distribution is found, as
in the case of ultracold atomic collisions. Interestingly, this
implies that effects of anisotropy are to be seen at the lowest
temperatures only when the scattering length is small, and cross
sections are accurately represented by the Born approximation.

\section{What this means for you}

Scaled units are fine for proving a theoretical point, as we have
hoped to do here. However, since dipole length scales vary widely
between different molecules and at different electric field
strengths, it is also useful to consider specific examples that
measure cross sections in cm$^2$.

Before presenting such an example, we first recapitulate our main
results, cast in terms of the explicit dimensionful factors.  The
Born result, valid in the ultracold limit, is
\begin{eqnarray}
\sigma_{\rm Born}^e &=& 1.117 {M^2 \langle \mu_1 \rangle^2 \langle
\mu_2 \rangle^2 \over \hbar^4} + 4 \pi a^2 \nonumber \\
\sigma_{\rm Born}^o &=& 3.351 {M^2 \langle \mu_1 \rangle^2 \langle
\mu_2 \rangle^2 \over \hbar^4}.
\end{eqnarray}
The semiclassical result, valid for cold collisions, $E>E_D =
\hbar^6 / M^3 \langle \mu_1 \rangle^2 \langle \mu_2 \rangle^2$, is
\begin{eqnarray}
\sigma_{\rm Ei} = {8 \pi \over 3} { \langle \mu_1 \rangle \langle
\mu_2 \rangle \over \hbar} \sqrt{ {M \over 2 E}}.
\end{eqnarray}
(Recall that our eikonal derivation does not distinguish between
even and odd partial wave contributions.  To a good approximation,
both the even and odd contributions would be half this value.) These
formulas show explicitly that the cross sections at low and high
energy differ not only in their dependence on energy, but also in
their dependence on the parameters -- reduced mass and dipole
moments -- of the molecules.

\begin{figure}
\includegraphics[width=0.9\textwidth] {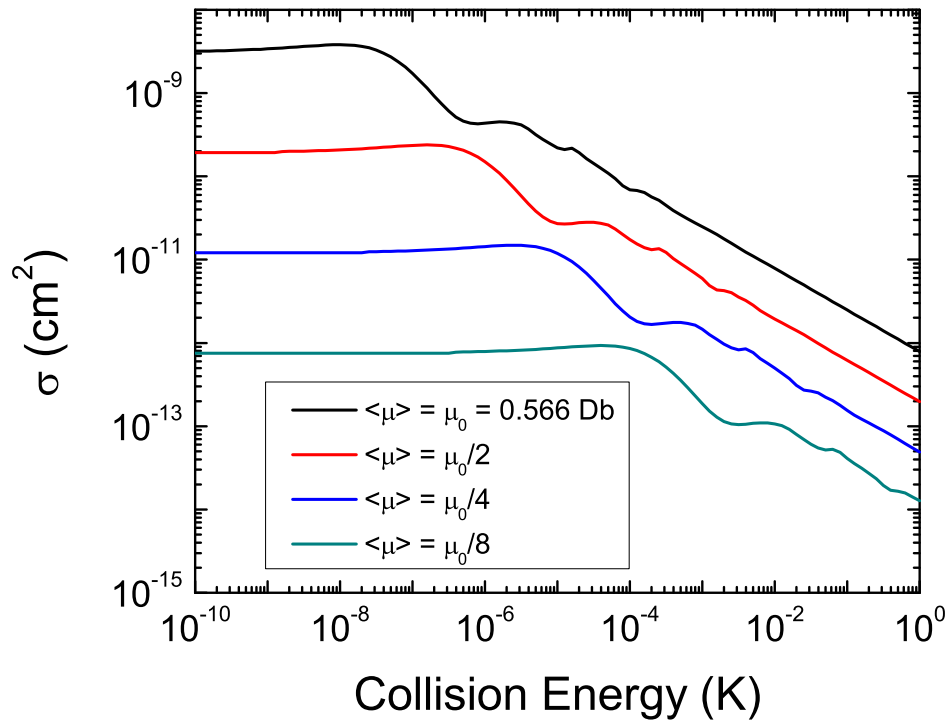} \caption{Elastic cross section for
scattering of pairs of fermionic $^{40}$K$^{87}$Rb molecules in
identical internal states, averaged over incident directions.  This
calculation is based on the ``universal'' calculation that includes
only dipole-dipole interactions. The top curve is the cross section
for fully polarized molecules with dipole moment $\langle \mu
\rangle = 0.566$ Debye. The dipole is halved for each successively
lower curve.}
\end{figure}

To give a concrete example, consider the ground-state
$^{40}$K$^{87}$Rb molecules that were recently produced at
temperatures of several hundred nK \cite{gpm}.  This molecule has a
dipole moment of 0.566 Debye, and, being a fermion, would collide
only in odd partial waves if it is trapped in a single quantum
state. We therefore plot in Fig. 5 the odd partial wave cross
section computed above, but cast in realistic units for this
molecule.  The largest cross section corresponds to the full dipole
moment, $\langle \mu \rangle = 0.566$ Debye.  Each successively
lower curve divides the dipole moment in half from the previous one.
For this reason, each low-energy cross section drops by a factor of
16 from the one above, while at high energy each cross section drops
by a factor of 4. Because the dipole moment is something that can be
changed by the application of a greater or lesser electric field,
cross sections spanning this stunning range of magnitudes should be
observable in experiments.

Also interesting is the energy scale encompassed by this figure. In
the experiment, the gas is trapped at a temperature of 350 nK.  At
low electric field values, hence low dipole moments, these molecules
are in the ultracold regime, and scatter according to the Born
prescription.  At higher fields, however, $E_D$ approaches the
temperature of the gas, and experiments might start to observe the
non-universal behavior of the scattering.

In summary, we have characterized the total scattering cross section
for dipolar molecules, both in the cold limit $E_D < E <B_e$, and in
the ultracold limit $E<E_D$.  The behavior of this scattering is
universal for cold collisions, and nearly so for ultracold
collisions.  In the temperature regime intermediate between these
two, universality breaks down, and empirical cross sections will
likely reveal information about the intermolecular potential energy
surface.

The authors acknowledge the financial support from the  NSF (J. L.
B.), the ARC (C. T.), and the University of Kentucky (M. C.)

\end{document}